\documentclass{PoS}

\newcommand{\nn}{\nonumber}
\newcommand{\ber}{\begin{eqnarray}}
\newcommand{\eer}{\end{eqnarray}}
\newcommand{\pa}{\partial}
\newcommand{\bbD}[1]{\mathbb{D}_{#1}}
\newcommand{\bbDB}[1]{\bar{\mathbb{D}}_{#1}}
\def\+{{+\!\!\!+}}
\def\one{1\!\!\!1}
\def\pp{\mbox{\tiny${}_{\stackrel\+ =}$}}

\title{Uses of Sigma Models}

\ShortTitle{Uses of Sigma Models}

\author{\speaker{Ulf Lindstr\"om}%
       \thanks{Imperial-TP-2018-UL-01, Uppsala UUITP-08/18. The title of this presentation is a paraphrase on the title of   \cite{Coleman:1978ae}. It gives me a chance to cite this great article by Sidney Coleman, no other comparison intended.}\\
      Uppsala, SE and Imperial, UK\\
      E-mail: \email{ulf.lindstrom@physics.uu.se}}

\abstract{This is a brief review of some of the uses of nonlinear sigma models. After a short general discussion touching on point particles, strings and condensed matter systems, focus is shifted to sigma models as probes of target space geometries. The relation of supersymmetric non-linear sigma models to K\"ahler, hyperk\"ahler, hyperk\"ahler with torsion and generalised K\"ahler geometries is described.}

\FullConference{Corfu Summer Institute 2017 "School and Workshops on Elementary Particle Physics and Gravity"\\
		 2-28 September 2017\\
		 Corfu, Greece}

\begin{document}
\section{Introduction}
Sigma models are maps from a domain $\Sigma$ , taken to be a (super) manifold, into another (super) manifold $\cal T$ called the target space. Mathematically these models may be described as ``harmonic maps''  due to their geometric target space interpretation, whereas they are typically thought of as lower dimensional field theories from the point of view of $\Sigma$.

Historically the  name derives from an article by Gell-Mann and Levy \cite{GellMann:1960np} who introduced  both a particular linear and a particular non-linear model for to describe a particle $\sigma$ participating in beta decay together with pions.

The general (bosonic) nonlinear sigma model with its geometric interpretation was gradually introduced in the late 1970s. See, e.g.,  \cite{NLSM}.

In this very brief and biased review, we shall take a quick look at some of the areas where sigma models play a role and then focus on their importance as probes of the geometry on $\cal T$, in particular when combined with supersymmetry. This latter field was pioneered by Zumino in  \cite{Zumino:1979et} and by Alvarez-Gaume and Freedman in \cite{AlvarezGaume:1980vs}.

Nonlinear sigma models show perturbative renormalisability and asymptotic freedom \cite{Blasi:1989dj}, \cite{Anagnostopoulos:2010gw}.
 They have a number of supersymmetric \cite{Zumino:1979et}, \cite{AlvarezGaume:1980vs}, \cite{Gates:1984nk}, \cite{Hull:1985jv},
as well as  Wess-Zumino-Witten extensions \cite{Wess:1971yu}, \cite{Witten:1983tw}, and describe phenomenon ranging from spontaneous symmetry breaking \cite{GellMann:1960np} to
the dynamics of string world-sheets  \cite{Scherk:1974jj}. Below, in sec. $\!3$, are some examples.

\section{Definition}
The maps from $\Sigma$ to $\cal T$ we considered are
\ber
\phi^i:\Sigma\to{\cal T}~:~~~x^\mu \mapsto \phi^i(x) 
\eer
where $x^\mu, ~\mu=1,...,d,$ are coordinates on $\Sigma$ and their images $\phi^i(x),~ i=1,...,n,$  coordinates on $\cal T$. The actual form of the maps is found by extremizing the action
\ber\label{def}
S=\int dx \left(g_{ij}(\phi) \partial_\mu \phi^i\partial^\mu \phi^j+\dots\right)~, 
\eer
where the symbolic measure $dx$ and possible additional terms will depend on the dimension of $\Sigma$, the number of supersymmetries etc.  Extremizing $S$ with respect to $\phi$ leads to the equation
\ber
g_{ij}(\phi)\left( \partial^2 \phi^j+\partial^\mu\phi^k\Gamma_{kn}^{~~j}\partial_\mu\phi^n+\dots \right)=0
\eer
The symmetric tensor $g_{ij}$ is a metric on $\cal T$, and $\Gamma$ is is a metric connection, (in $2d$ possibly including torsion). Thus interpreted, the equation is the pullback of the covariant Laplacian, hence the name ``harmonic map''.

When (\ref{def}) is extended to include a potential for the scalar fields it is called a Landau Ginsberg model and has important applications in solid state theory \cite{Ginzburg:1950sr}.

\section{Examples}
To be explicit, here are a few examples. The subscript on $\Sigma$  indicates its dimension:
\bigskip

$\bullet~~$  Quantum Mechanics: $ ~~~~~\Sigma_1=$ time, ~~${\cal T}=$ line 
\bigskip

$\bullet~~$ Relativistic particle: $~~~~~~~\Sigma_1=$ time, ~~${\cal T}=$ Spacetime,~ $S=\int dx^0 \left(g_{ij}(X) \dot X^i\dot X^j\right) $
\bigskip

$\bullet ~~$ Relativistic string (in conformal gauge):
 $~~~\Sigma_2=(\tau,\sigma), ~~{\cal T}=$ Spacetime 
 \bigskip 
 
$\bullet~~$  $2d$ sigma model:$~~~~~\Sigma_2, ~~{\cal T}=$ General,~~
S=$\mu\int d^2x \left(g_{ij}(\phi) \eta^{ab}\partial_a\phi^i\partial_b \phi^j+B_{ij}(\phi)\epsilon^{ab}\partial_a\phi^i\partial_b \phi^j\right) $
\bigskip

We note that  $\mu$ is a dimensionless coupling. Correspondingly, classically the model is conformally invariant.  As a two dimensional field theory, the metric and $B$ field should be interpreted as generalised coupling constants. In the quantum theory they will in general scale with energy.
For the case of vanishing $B$ field, Friedan \cite{Friedan:1980jm} showed  that it obeys a renormalisation group equation which may be written
\ber
\lambda \frac{\partial g_{ij}}{\partial \lambda}=\beta_{ij}(\mu g)=R_{ij} + O(\mu^2)
\eer
where $ \lambda^{-1} $ is a short distance  cut off. This equation links the renormalisation group to Ricci flow: The one-loop renormalisation group flow for the nonlinear sigma model
equals the Ricci flow on the target manifold. Solutions to the Einstein's equation are  thus fixed points.

\bigskip

$\bullet$  The  $O(3)$ model:~~~~~$\Sigma_2, ~~{\cal T}= S^2,~~~~S=\int d^2x (\phi) \eta^{ab}\partial_a\hat n\partial_b \hat n :~~~\hat n^2=1$
\bigskip

For this model, the finite action solutions are instantons \cite{Coleman:1978ae} $S^2\to S^2$ classified by the second homotopy group of $S^2$. 
\bigskip

 $\bullet$   The  $O(n)$ model: ~~~~~$\Sigma_{d}, ~~ i,j=1,...,n=dim~\!{\cal T}$, ~~~~$S=\mu\int d^dx g_{ij}(\phi)\partial^\mu\phi^i\partial_\mu \phi^ j -V(\phi)$
\bigskip

The O(n) model is an effective field theory that encodes the pattern of symmetry breaking of an O(N) classical Heisenberg model defined on a square lattice. This model has non trivial critical behaviour in spite of $\mu$ being dimensionful.  It has applications in solid state theory.
\bigskip

$\bullet $ The  $G/H$ coset models:~~~~~$\Sigma_{d}, ~~{\cal T}=G/H$,  $G$ Lie group, $G\supset H$.
\bigskip

These models arise in the scalar sector of supergravity theories \cite{Freedman:2012zz}. As effective field theory for scalars it has the advantage of avoiding relevant operators.
\bigskip

$\bullet$ General $d$-dimensional model.~~~$\Sigma_{d}, ~~{\cal T}$,   ~~The number of supersymmetries dictate the geometry of ${\cal T}$. 
\bigskip

The target space geometry of $4d$ models with one supersymmetry was first identified by Zumino, \cite{Zumino:1979et} and extended to two supersymmetries in  \cite{AlvarezGaume:1980vs}. In $2d$ where the target space geometry can have torsion  $(2,2)$ and  $(4,4)$ supersymmetry was discussed  in \cite{Gates:1984nk} and  $(p,q)$ supersymmetry in  \cite{Hull:1985jv}. The  situation is summarised in the following tables
\eject 

 \begin{table}[!h]
 \begin{center}

  \begin{tabular}[H]{|l|lll|l| }
  \hline
  d = &6&4&2&Geometry\cr
  \hline
  ${\cal N}$ = &1&2&4&Hyperk\"ahler\cr
  ${\cal N}$ = &~&1&2&K\"ahler\cr
  ${\cal N}$ = &~&~&1&Riemannian\cr
  \hline
  \end{tabular}
  \caption{}
    \end{center}
  \end{table}

 Odd dimensions have the same structure as the even dimension lower. When we specialise to two or six dimensions, we have the additional possibility of having independent left and right supersymmetries.\bigskip

\noindent
In
$2d$ with $p=q$:
\normalsize
\begin{table}[!h]
\centering
\begin{tabular}{|l|c|c|c|c|c|}
\hline
Susy &(1,1) & (2,2) &{(2,2)}& (4,4) & (4,4) \\ 
\hline
E=g+B & $g,B$ &  $g$ & ${g,B}$ & $g$ & $g,B$ \\
\hline
Geometry  &  Riemannian &K\"ahler & Bihermitean  &Hyperk\"ahler &Bihypercomplex\\
\hline
\end{tabular}
 \caption{}
\end{table}

For the cases with $p\ne q$ typically the geometry  characterising the whole tangent bundle of ${\cal T}$ is determined, as discussed in \cite{Hull:1985jv} and in, e.g., \cite{Dine:1986by} and \cite{Howe:1988cj}.

\section{Supersymmetry}

The relations in Tables 1 and 2 are most easily derived in superspace formulations of the sigma models. 
The  $(2,2)$ geometry in $2d$ of follows from the dimensional reduction the  ${\cal N}=1$ in $4d$ . Using Weyl spinors the $4d$ scalar superfields and covariant derivatives are the extensions to superspace of the scalar fields and derivatives according to
 \ber\nn
 &&\phi(x)\to\phi(x,\theta,\bar\theta)\\[1mm]
 &&\partial_\mu\to(\partial_\mu, D_\alpha,\bar D_{\dot \alpha})~,
 \eer
 where $\theta^\alpha$ are anticommuting spinorial (odd) coordinates. The covariant derivatives obey
 \ber
 \{D_\alpha,\bar D_{\dot \alpha}\}=2i(\sigma^\mu)_{\alpha\dot\alpha}\partial_\mu=:2i\partial_{\alpha\dot\alpha}~,
 \eer
 and a scalar superfield has a finite expansion in the odd coordinates
 \ber
 \phi^i=\phi^i(x)+\theta^\alpha\psi_\alpha(x)+...+\theta^\alpha\bar\theta^{\dot\alpha}
 v^i_{\alpha\dot\alpha}(x)+..  ~.
 \eer
 An irreducible representation of supersymmetry is obtained by imposing a chirality constraint:
 \ber\label{exp}
\bar D_{\dot \alpha}\phi=0~,~~\Rightarrow ~~~\phi=\phi+\theta^\alpha\psi_\alpha+\theta^\alpha\theta_\alpha{\cal F}~,
 \eer{}
 reducing the multiplet to one complex scalar $\phi$, one Weyl spinor $\psi_\alpha$ and one complex pseudo scalar ${\cal F}$ in terms of Minkowski space fields.
 The expansion (\ref{exp}) has this simple form in a representation where $\bar D_{\dot\alpha}= \pa / {\pa\bar\theta^{\dot \alpha}}$ but $D_\alpha$ is more complicated. A representation-independent way of defining the component fields is
 \ber
 \phi(x)=\phi(x,\theta,\bar\theta)_|~,~~~\psi_\alpha(x)=D_\alpha\phi(x,\theta,\bar\theta)_|~,~~~{\cal F}(x)={\textstyle \frac 1 2}D^\alpha D_\alpha\psi_\alpha(x,\theta,\bar\theta)_|~,
 \eer
 where a vertical bar denotes setting the odd coordinates to zero. This also makes the derivation of component actions easy.  
 
 \section{Supersymmetric sigma model geometry}
 
 The most general superspace action for $n$ chiral fields $\phi^i, i=1,...,n$  is given by a real  function $K$ and may  be evaluated as
 \ber
 S=\int d^4x D^2\bar D^2K(\phi,\bar\phi)_|=\int d^4x \left\{{\color{red}K_{i\!~\bar j}}\partial_\mu\phi^i\partial^\mu\bar\phi^{\bar j}+....\right\}~,
 \eer
 where the term on the right only involves the lowest components of the superfield and the dots denote supersymmetric completion. Comparing to the definition of a sigma model (\ref{def}) we conclude that this is a supersymmetrisation of a model with a complex target space geometry where the metric has a potential $g_{i\bar j}={\pa^2 K}/{\pa\phi^i\bar\pa\phi^{\bar j}}.$
 This is K\"ahler  geometry, the $4d$ Table 1 entry for ${\cal N}=1$ and one of the $2d$ Table 2 entries for $(2,2)$.
 
The  $(4,4)$ models in $2d$  are reductions of  ${\cal N}=2$ models  in $4d$. For these, 
 the action is as as for the K\"ahler models above, but with extra, non-manifest supersymmetry: 
 \ber\label{extra}
\delta\phi^i =\bar D^2(\bar\varepsilon\bar\Omega^i)~,\quad \delta \bar\phi^{\bar i} =D^2(\varepsilon\Omega^{\bar i})~,
\eer{}
where $\Omega=\Omega(\phi,\bar\phi)$. Its derivatives may be combined into \cite{Hull:1985pq}
\ber\label{nonman}
J^{(1)}=\left(\begin{array}{cc}
0&\Omega^{\bar i}_{~j}\\
\bar\Omega^i_{~\bar j} &0\end{array}\right)\qquad 
J^{(2)}=\left(\begin{array}{cc}
0&i\Omega^{\bar i}_{~j}\\
-i\bar\Omega^i_{~\bar j} &0\end{array}\right)
\eer{}
which are considered together with the fundamental complex structure used to define $\phi$ and $\bar\phi$ as complex
\ber\label{can}
J^{(3)}=\left(\begin{array}{cc}
i\delta^i_j&0\\
0 &-i\delta^{\bar i}_{\bar j}\end{array}\right)
\eer{}

Invariance of the action under the transformations (\ref{extra}) as well as closure of the algebra will follow iff the  $J^{(A)}$s  are complex structures with
$J^{(A)}J^{(B)}=-\delta^{AB}+\epsilon^{ABC}J^{(C)}$, i.e., they satisfy an $SU(2)$ algebra. This is hyperk\"ahler geometry, the $4d$ Table 1 entry for ${\cal N}=2$ and one of the $2d$ Table 2 entries for $(4,4)$.

\subsection{$4d$ models: K\"ahler and hyperk\"ahler geometry}
Symplectic manifolds with symmetries lend themselves to a reduction, or quotient, as described in \cite{MarsWein}, \cite{GuilleStern}. The key issue here is to preserve properties such being symplectic. A K\"ahler manifold is symplectic with extra structure, and so is its quotient provided that it is arrived at by gauging holomorphic Killing symmetries. This is the K\"ahler quotient which may be used to find new K\"ahler spaces. 

As we saw above, finding a new K\"ahler space is equivalent to finding a new Lagrangian $K$. To find a new hyperk\"ahler geometry starting from an old one, it is necessary to preserve much more structure. For certain isometries, this can be done guided by the requirement that the extended supersymmetry be preserved. In brief the recipe for this hyperk\"ahler  reduction is as follows \cite{Lindstrom:1983rt}, \cite{Hitchin:1986ea} 

\begin{itemize}
\item The starting point is  a hyperk\"ahler potential with triholomorphic isometries, i.e., isometries that are holomorphic with respect to all three $J^{(A)}$s.
\item These isometries are gauged in preserving the supersymmetries, as described in \cite{Hull:1985pq}.
\item The resulting action is extremised with respect to the ${\cal N}=2$ multiplet of gauge fields. This will result in a particular form of the vector potential plus certain constraints. This all determines the form of the extremised action as a ${\cal N}=2$ sigma model on the quotient space.
\item From this action, we read off the resulting quotient metric and complex structure of the new hyperk\"ahler  manifold.
\end{itemize}
 The hyperk\"ahler reduction is particularly useful since it gives a way of finding new hyperk\"ahler geometries which, unlike in the K\"ahler case, cannot be done simply by writing down a new action.\footnote{Technically, this is because there is no off shell $(4,4)$ action formulation (not involving additional variables) available.}
 \subsection{$2d$ models: Generalised K\"ahler geometry}
 As mentioned, some of the $(2,2)$ and $(4,4)$ models in $2d$ are dimensionally reduced $4d$ models. However, there are new and interesting  possibilities that arise in $2d$.

Spinor and vector indices in $2d$ take on two values only, which we denote $(+,-)$ and $(\+,=)$ respectively. The $(2,2)$ superspace thus has coordinates $(x^\+,x^=,\theta^+, \theta^-)$ and the $(2,2)$ supersymmetry algebra is
\ber
\{\bbD{\pm},\bbDB{\pm}\}=i\partial_{\pp}~.
\eer
The spinorial covariant derivatives $\bbD{}$ may be used to impose a number of covariant constraints on the $(2,2)$ superfields:
\ber\nn
\bbDB{\pm}\phi=0~,\\[1mm]\nn
\bbDB{+}\chi=\bbD{-}\chi=0~,\\[1mm]\nn
\bbDB{+}\ell=0~,\\[1mm]
\bbDB{-}r=0~,
\eer{}
along with their complex conjugates. These define Chiral ${\tilde\phi}:=(\phi,\bar\phi)$, Twisted Chiral $\tilde\chi:=(\chi,\bar\chi)$ \cite{Gates:1984nk}, Left semichiral $L:=(\ell,\bar\ell)$ and Right semichiral $R:=(r,\bar r)$ superfields \cite{Buscher:1987uw}.
A general $(2,2)$ action in $2d$ involves a function $K$:
\begin{equation}\label{GKG}
S=\int d^2x\bbD{+}\bbDB{+}\bbD{-}\bbDB{-}K(\tilde\phi,\tilde\chi,L,R)
\end{equation}
For this function to describe a sigma model it has to involve an equal number of left and right semichiral fields and satisfy some regularity conditions. The target space geometry may be described as bihermitean geometry \cite{Gates:1984nk}, $(g,J^{(\pm)}, H)$ i.e., with a metric $g$  hermitean with respect to two complex structures $J^{(\pm)}$  covariantly constant with respect to  connections with torsion determined by a closed three form  $ H$. This is a special case of Generalised Complex Geometry  \cite{Hitchin:2004ut}, a geometry defined on the sum of the tangent and cotangent bundles of a manifold. In this formulation, bihermitean geometry corresponds to Generalised K\"ahler geometry \cite{Gualtieri:2003dx}.
In \cite{Lindstrom:2005zr} it is shown that the inclusion of semichiral fields in the function $K$ in (\ref{GKG}) gives a full description of Generalized K\"ahler geometry, away from certain irregular points. 

The function $K(\tilde\phi,\tilde \chi, L, R)$ has several additional roles discovered from the sigma model point of view: It is  is the superspace Lagrangian for a $(2,2)$ sigma model with
 Generalized K\"ahler target space geometry. It is also the generalized K\"ahler potential for the metric $g$ and $B$-field. These fields are determined as (non-linear) functions of the Hessian of $K$. In addition, $K$ generates symplectomorphisms between Darboux coordinates for $J^{(+)}$ and Darboux coordinates for  $J^{(-)}$ \cite{Lindstrom:2005zr}.
 
 \subsection{$2d$ models: hyperk\"ahler geometry with torsion}
 \subsubsection{($4,1)$ in $(2,1)$ superspace}

The $2d$ Table 2 does not contain the general
$(p,q)$ models introduced in \cite{Hull:1985jv}, \cite{Hull:1986hn}. Some  of the geometry was extensively discussed in \cite{Dine:1986by} and in $(1,0)$ superspace in \cite{Howe:1988cj}. The number of left and right supersymmetries, denoted {\em (left,right)=(p,q)},  restrict the target space geometry $(g,H=dB,J^{(A)},...)$ to be  (hyper)k\"ahler with torsion (typically). As an illustration, we will describe  two recent discussions of $(4,q)$ models \cite{Hull:2016khc}, \cite{Hull:2017hfa} .

The superspace is $(2,1)$ superspace with supersymmetry algebra
\ber
\{\bbD{+},\bbDB{+}\}=i\partial_{\+}~,~~~D_-^2=i\partial_{=}~,
\eer
and the action is
\ber\label{2,1}
S= 
\int d^2x d^3 \theta
\left(k_ i D_-\varphi^i
+
\bar k_{\bar i}  D_-\bar \varphi^{\bar i} 
\right)~.
\eer{}
The fields  $\varphi^i, i=1,...,n$ are $(2,1)$ chiral
\ber
\bbDB{+}\varphi^i=0~,
\eer{}
and  $\bar \varphi^{\bar i} =
(\varphi^i)^*$. The metric and $B$ field are determined by the complex vector potentials as 
\ber\nn\label{gB}
&&g_{i \bar j}=i(\pa_{ i} \bar k_{\bar  j}-\pa_{\bar j} k_{i} )           \\[1mm]\nn
&&B^{(2,0)}_{ij}=  i(  
\pa_{i} k_j     -\pa_{ j}k_{i }  )
\\[1mm]
&&B=B^{(2,0)}+B^{(0,2)}
\eer

\bigskip

An ansatz for the nonmanifest $(4,1)$ Susy is:

\ber\nn
\label{fvar}
&&\delta\varphi^i =\bar\epsilon^+\bbDB{+}\Omega^i(\varphi,\bar\varphi)\\[1mm]
&&\delta\bar\varphi^{\bar i} = \epsilon^+\bbD{+}\bar \Omega^{\bar i}(\varphi,\bar\varphi)
\eer
This is formally the same ansatz as for the ${\cal N}=2$ models  in $4d$, given in (\ref{extra}), and leads to the same set of complex structures (\ref{nonman}), (\ref{can}) forming an $SU(2)$ algebra for the left sector, albeit the complex structures are now covariantly constant with respect to the torsionful connection
\ber
\Gamma^{(+)}:=\Gamma^{(0)}+\textstyle \frac 1 2 g^{-1}H~,~~~H=dB~,
\eer{}
with $\Gamma^{(0)}$ the Levi-Civita connection. The geometry is thus hyperk\"ahler with torsion.

We can find an off-shell $(4,1)$ multiplet that realises the geometry. The $(4,1)$ algebra is
\ber\nn\label{talg1}
&&\{\bbD{+a},\bbDB{+} ^b\}=~2i\delta^b_a\pa_\+~, ~~~a,b,=1,2.\\[1mm]
&&(D_-)^2=i\pa_=~.
\eer{}
 The $(4,1)$ multiplets  we study are $\varphi^i=(\phi^i,\chi^i)$ with
 \ber\nn\label{constr2}
&&\bbDB{+}^1\phi = 0=\bbD{+2}\phi~,~~~\bbDB{+}^1 \chi =0=\bbD{+2}\chi~,\\[1mm]
&&\bbDB{+}^2\chi=-i\bbDB{+}^1\bar \phi~,~~~\bbDB{+}^2\phi=i\bbDB{+}^1\bar\chi~.
\eer{}

These are off shell multiplets and give a geometry with
\ber\label{comstr}
\mathbb{J}^{(A)}= \mathbb{I}^{(A)}\otimes \one_{n\times n}
\eer{}
where
\ber\label{comstr1}
\mathbb{I}^{(1)}=\left(\begin{array}{cc}i\one&0\\
0&-i\one\end{array}\right)~,~~
\mathbb{I}^{(2)}=\left(\begin{array}{cc}0& i\sigma_2\\ i\sigma_2&0\end{array}\right)
~,~
\mathbb{I}^{(3)}=\left(\begin{array}{cc}0&-\sigma_2\\
\sigma_2&0\end{array}\right)~.
\eer
with $\sigma_2$ the second of the Pauli matrices.

The metric and torsion arise from an action. A convenient formulation is via projective superspace, an extension of superspace by an additional coordinate 
 $\zeta\in \mathbb{CP}^1$, introduced for ${\cal N}=2$ in $4d$ in \cite{Karlhede:1984vr}. It may be adapted to the present case introducing
 \ber\nn
&&\nabla_+:=\bbD{+1}+\zeta\bbD{+2}~,\\[1mm]\nn
&&\breve{\nabla}_+:=\bbDB{+}^1-\zeta^{-1}\bbDB{+}^2~,\\[1mm]
&&\eta=\sum \eta_m\zeta^m~,~~~~~\nabla_+\eta=\breve{\nabla}_+\eta=0~,
\eer{}
where $\eta$ is called a projectively chiral superfield. As an invariant action we may take
\ber\label{akt}
S=i\int d^2x
\oint_C\frac{d\zeta}{2\pi i \zeta}\bbD{+}\bbDB{+}D_-\left(\lambda_i(\eta,\breve{\eta};\zeta)D_-\eta^i-\breve{\lambda}_i(\eta, \breve{\eta};\zeta)D_-\breve{\eta}^i\right).
\eer{}

For the multiplets introduced in (\ref{constr2}) we have
\ber\label{simpeta}\nn
&&\eta^i=\bar\phi^i+\zeta \chi^i\\[1mm]
&&\breve{\eta}^i=\phi^i-\zeta^{-1}\bar\chi^i~.
\eer{}
From the action (\ref{akt})we read off the one form potential
\ber\nn\label{ks}
&&k_{\phi^i}=-\oint_C\frac{d\zeta}{2\pi i \zeta}\breve{\lambda}_i~,~~~\bar k_{\bar\phi^i}=\oint_C\frac{d\zeta}{2\pi i \zeta}{\lambda}_i\\[1mm]
&&k_{\chi^i}=\oint_C\frac{d\zeta}{2\pi i \zeta}\zeta{\lambda}_i~,~~~~\bar k_{\bar\chi^i}=\oint_C\frac{d\zeta}{2\pi i \zeta}\zeta^{-1}\breve{\lambda}_i~.
\eer{}
Using  these we may check that the geometry is indeed hyperk\"ahler with torsion.

 \subsubsection{($4,0)$ in $(2,0)$ superspace}
 Finally, we briefly comment on yet one more off shell model: An off shell realisation of a $(4,0)$ sigma model.
 
 The $(4,0)$ action in $(2,0)$ superspace is
\ber\label{2,0}\nn
&&S= \int d^2x d^2 \theta
\left(k_ i \partial_=\varphi^i
+
\bar k_{\bar i}  \partial_= \bar \varphi^{\bar i}
+e_{\mu \nu}\Lambda_-^{ \mu }\Lambda_-^{ \nu} + {G_{\mu \bar\nu}}\Lambda_-^{ \mu }\bar\Lambda_-^{\bar \nu}
+e_{\bar\mu \bar\nu}\bar\Lambda_-^{ \bar \mu }\bar\Lambda_-^{\bar \nu}
\right)~,
\eer{}
The $(2,0)$ chiral scalar  and fermion superfields satisfy,
\ber\nn
\bbDB{+}\varphi^i=0~,~~~\bbDB{+}\Lambda_-^{ \mu }=0~.
\eer{}
The metric $g$ and $B$ field on the target space ${\cal T}$ are given in terms of the vector potentials as in (\ref{gB}).
The model is formulated on a bundle space with base  ${\cal T}$ ,  with $G_{\mu\bar\nu}$  a fibre metric and  $e_{\mu\nu}$ an antisymmetric field, both related to a bundle connection $A_i{}^M{}_N$.
The conditions for $(4,0)$ supersymmetry include the existence of an $SU(2)$ worth of complex structures on this bundle (as well as on the base and fibre spaces). Again there is an off shell  realisation of the geometry \cite{Hull:2017hfa}.

\section{Closing comments}

This brief presentation  only touches on some of the areas where sigma models have found applications. The format has made it necessary to focus on one particular aspect when going into more detail; the target space geometry. Many other aspects are not treated, such as quantum aspects, $T$-duality, topological sigma models, the relation to string vacua, to gravity and supergravity solutions and the precise relation to spin models. It is hoped, however, that it is conveyed that Sigma Models are versatile tools that have a large number of uses.
 \bigskip
 
\noindent {\bf Acknowledgement:}\\
 I am grateful to all my collaborators over the years on this topic, in particular to Martin Ro\v cek who initiated our research into sigma mode geometry all those years ago.
 I  also gratefully acknowledge the hospitality of the theory group at Imperial College, London, as well as support from the EPSRC programme grant ''New
Geometric Structures from String Theory'' EP/K034456/1.


\begin{thebibliography}{99}
\bibitem{GellMann:1960np} 
  M.~Gell-Mann and M.~Levy,
  ``The axial vector current in beta decay,''
  Nuovo Cim.\  {\bf 16}, 705 (1960).
 
 
 \bibitem{NLSM}
  A.~M.~Polyakov,
  ``Interaction of Goldstone Particles in Two-Dimensions. Applications to Ferromagnets and Massive Yang-Mills Fields,''
  Phys.\ Lett.\  {\bf 59B}, 79 (1975).
  doi:10.1016/0370-2693(75)90161-6
  

  A.~M.~Polyakov and A.~A.~Belavin,
  ``Metastable States of Two-Dimensional Isotropic Ferromagnets,''
  JETP Lett.\  {\bf 22} (1975) 245
   [Pisma Zh.\ Eksp.\ Teor.\ Fiz.\  {\bf 22} (1975) 503].
  
  
  A.~Jevicki,
  ``Quantum Fluctuations of Pseudoparticles in the Nonlinear Sigma Model,''
  Nucl.\ Phys.\ B {\bf 127} (1977) 125.
  doi:10.1016/0550-3213(77)90355-8
  
  D.~Forster,
  ``On the Structure of Instanton Plasma in the Two-Dimensional O(3) Nonlinear Sigma Model,''
  Nucl.\ Phys.\ B {\bf 130}, 38 (1977).
  doi:10.1016/0550-3213(77)90391-1

C J Isham, in ``Relativity, Groups and Topology II'', BS DeWitt and R Stora, eds., North Holland, Amsterdam (1984).

\bibitem{Friedan:1980jm} 
  D.~H.~Friedan,
  ``Nonlinear Models in Two + Epsilon Dimensions,''
  Annals Phys.\  {\bf 163}, 318 (1985).


\bibitem{Zumino:1979et} 
  B.~Zumino,
  ``Supersymmetry and Kahler Manifolds,''
  Phys.\ Lett.\  {\bf 87B}, 203 (1979): Addendum,
  CERN-TH-2733-Add. (1979).
  
\bibitem{AlvarezGaume:1980vs} 
  L.~Alvarez-Gaume and D.~Z.~Freedman,
  ``Ricci Flat Kahler Manifolds and Supersymmetry,''
  Phys.\ Lett.\  {\bf 94B}, 171 (1980).
  
\bibitem{Blasi:1989dj} 
  A.~Blasi,
  ``Renormalization of nonlinear sigma models: A short review,''
  Nucl.\ Phys.\ Proc.\ Suppl.\  {\bf 16}, 574 (1990).
  
\bibitem{Anagnostopoulos:2010gw} 
  K.~Anagnostopoulos, K.~Farakos, P.~Pasipoularides and A.~Tsapalis,
  ``Non-Linear Sigma Model and asymptotic freedom at the Lifshitz point,''
  arXiv:1007.0355 [hep-th].
  
\bibitem{Gates:1984nk} 
  S.~J.~Gates, Jr., C.~M.~Hull and M.~Ro\v cek,
  ``Twisted Multiplets and New Supersymmetric Nonlinear Sigma Models,''
  Nucl.\ Phys.\ B {\bf 248}, 157 (1984).
  
\bibitem{Hull:1985jv} 
  C.~M.~Hull and E.~Witten,
  ``Supersymmetric Sigma Models and the Heterotic String,''
  Phys.\ Lett.\  {\bf 160B}, 398 (1985).
  
\bibitem{Wess:1971yu}
  J.~Wess and B.~Zumino,
  ``Consequences of anomalous Ward identities,''
  Phys.\ Lett.\  {\bf 37B} (1971) 95.
  doi:10.1016/0370-2693(71)90582-X
  
\bibitem{Witten:1983tw}
  E.~Witten,
  ``Global Aspects of Current Algebra,''
  Nucl.\ Phys.\ B {\bf 223} (1983) 422.
  doi:10.1016/0550-3213(83)90063-9

\bibitem{Scherk:1974jj}
  J.~Scherk,
  ``An Introduction to the Theory of Dual Models and Strings,''
  Rev.\ Mod.\ Phys.\  {\bf 47} (1975) 123.
  doi:10.1103/RevModPhys.47.123
  
  
\bibitem{Ginzburg:1950sr}
  V.~L.~Ginzburg and L.~D.~Landau,
  ``On the Theory of superconductivity,''
  Zh.\ Eksp.\ Teor.\ Fiz.\  {\bf 20} (1950) 1064.
  
\bibitem{Freedman:2012zz}
  D.~Z.~Freedman and A.~Van Proeyen,
  ``Supergravity,''
  
\bibitem{Hull:1986hn}
  C.~M.~Hull,
  ``Lectures On Nonlinear Sigma Models And Strings,''
  Lectures given at  the
Vancouver  Advanced Research Workshop, published in  {\it Super  Field
Theories}    (Plenum,  New  York,  1988),  edited  by  H.Lee   and
G.Kunstatter.
  
\bibitem{Dine:1986by} 
  M.~Dine and N.~Seiberg,
  ``(2,0) Superspace,''
  Phys.\ Lett.\ B {\bf 180}, 364 (1986).
  
  
\bibitem{Howe:1988cj} 
  P.~S.~Howe and G.~Papadopoulos,
  ``Further Remarks on the Geometry of Two-dimensional Nonlinear $\sigma$ Models,''
  Class.\ Quant.\ Grav.\  {\bf 5}, 1647 (1988).
  
\bibitem{Hull:1985pq} 
  C.~M.~Hull, A.~Karlhede, U.~Lindstr\"om and M.~Ro\v cek,
  ``Nonlinear $\sigma$ Models and Their Gauging in and Out of Superspace,''
  Nucl.\ Phys.\ B {\bf 266}, 1 (1986).
  
 \bibitem{MarsWein}
 J. ~Marsden and A.~ Weinstein, 
   ``Reduction of symplectic manifolds with symmetry'', 
 Rep.\ Math.\ Phys\ {\bf 5}, 121-130, (1974),

\bibitem{GuilleStern}
V.~Guillemin and S.~Sternberg,
 \underline{Symplectic techniques in physics}, Cambridge University Press 1984
 
\bibitem{Lindstrom:1983rt} 
  U.~Lindstr\"om and M.~Ro\v cek,
  ``Scalar Tensor Duality and N=1, N=2 Nonlinear Sigma Models,''
  Nucl.\ Phys.\ B {\bf 222}, 285 (1983).
  
\bibitem{Hitchin:1986ea} 
  N.~J.~Hitchin, A.~Karlhede, U.~Lindstr\"om and M.~Ro\v cek,
  ``Hyperkahler Metrics and Supersymmetry,''
  Commun.\ Math.\ Phys.\  {\bf 108}, 535 (1987).
  
\bibitem{Buscher:1987uw} 
  T.~Buscher, U.~Lindstr\"om and M.~Ro\v cek,
  ``New Supersymmetric $\sigma$ Models With {Wess-Zumino} Terms,''
  Phys.\ Lett.\ B {\bf 202}, 94 (1988).
  
  \bibitem{Hitchin:2004ut}
  N.~Hitchin,
  ``Generalized Calabi-Yau manifolds'', 
  Quart.\ J.\ Math.\ Oxford Ser.\  {\bf 54} (2003) 281
  [math/0209099 [math-dg]].  
  
  \bibitem{Gualtieri:2003dx}
  M.~Gualtieri, 
   ``Generalized complex geometry'', 
  math/0401221 [math-dg].
  
  \bibitem{Lindstrom:2005zr}
  U.~Lindstr\"om, M.~Ro\v cek, R.~von Unge and M.~Zabzine, 
  ``Generalized Kahler manifolds and off-shell supersymmetry'', 
  Commun.\ Math.\ Phys.\  {\bf 269} (2007) 833
  [hep-th/0512164].
  
\bibitem{Howe:1988cj}
  P.~S.~Howe and G.~Papadopoulos,
  ``Further Remarks on the Geometry of Two-dimensional Nonlinear $\sigma$ Models,''
  Class.\ Quant.\ Grav.\  {\bf 5} (1988) 1647.


  
\bibitem{Hull:2016khc} 
  C.~Hull and U.~Lindström,
  ``All $(4,1)$: Sigma Models with $(4,q)$ Off-Shell Supersymmetry,''
  JHEP {\bf 1703}, 042 (2017)
  doi:10.1007/JHEP03(2017)042
  [arXiv:1611.09884 [hep-th]].
  
  
\bibitem{Hull:2017hfa} 
  C.~Hull and U.~Lindström,
  ``All $(4,0)$: Sigma Models with $(4,0)$ Off-Shell Supersymmetry,''
  JHEP {\bf 1708}, 129 (2017)
  doi:10.1007/JHEP08(2017)129
  [arXiv:1707.01918 [hep-th]].
  
  \bibitem{Karlhede:1984vr} 
  A.~Karlhede, U.~Lindstr\"om and M.~Ro\v cek,
  ``Selfinteracting Tensor Multiplets in $N=2$ Superspace,''
  Phys.\ Lett.\ B {\bf 147}, 297 (1984).
  doi:10.1016/0370-2693(84)90120-5
  
  
\bibitem{Coleman:1978ae} 
  S.~R.~Coleman,
  ``The Uses of Instantons,''
  Subnucl.\ Ser.\  {\bf 15}, 805 (1979).
\end{thebibliography}
\end{document}